     \documentstyle[11pt]{article}
\font\titlfnt = cmss17

\begin{document}

\bibliographystyle{unsrt}

\renewcommand{\thesection}{\Roman{section}}
\renewcommand{\thefootnote}{\fnsymbol{footnote}}

\begin{titlepage}

\rightline{hep-th/9505189}\par

\vspace{7.5mm}

\begin{center}

{\titlfnt
Unique Closed-Form Quantization Via
Generalized Path Integrals or by
Natural Extension of the Standard
Canonical Recipe}

\vspace{10mm}

\noindent
{\large
S. K. Kauffmann}\footnote{
E-Mail address:
ken@catch22.com}\\
750 Gonzalez Drive, Apt.\ 6D, San Francisco, CA 94132-2208, U.S.A.

\vspace{5mm}

\end{center}

\vspace{10mm}

\begin{abstract}

The Feynman-Garrod path integral representation for
time evolution is extended to arbitrary
one-parameter continuous canonical transformations.
One thereupon obtains a generalized Kerner-Sutcliffe
formula for the unique quantum representation of the
transformation generator, which can be an arbitrary
classical dynamical variable.  This closed-form
quantization procedure is shown to be equivalent
to a natural extension of the standard canonical
quantization recipe---an extension that resolves
the operator-ordering ambiguity in favor of the
Born-Jordan rule.

\vspace{5mm}
\end{abstract}

\vspace{88mm}

\noindent
Revised November 1995

\end{titlepage}

\renewcommand{\thefootnote}{\arabic{footnote}}
\addtocounter{footnote}{-\value{footnote}}

%%%%%%%%%%%%%%%%%%%%%%%%%%%%%%%%%%%%%%%%%%%%%%%%%%%%%%%%%%%%%%%%%%%

\addtocounter{equation}{-\value{equation}}

The Feynman path integral \cite{Feynman}, as extended and
generalized by Garrod \cite{Garrod}, provides an explicit and
unique procedure for calculating the quantum theory time
evolution transformation in coordinate reprentation from its
purely classical canonical generator, the Hamiltonian function.
As a byproduct, the elegant Kerner-Sutcliffe formula for the
unique quantization of the Hamiltonian is obtained \cite{Kerner},
which turns out to resolve the operator-ordering ambiguity
in accord with the Born-Jordan rule \cite{Born,Cohen66}.
In this paper the Feynman-Garrod path integration procedure is
straightforwardly extended to the quantization of arbitrary
one-parameter continuous classical canonical transformations,
and a generalized Kerner-Sutcliffe formula is obtained for
the unique quantization of their generating functions, which
can be essentially arbitrary real-valued classical dynamical
variables.  This quantum representation is, of course, explicitly
Hermitian, and has the property that the quantum representation of
the Poisson bracket of a {\em function\/} of the canonical
coordinate with a {\em function\/} of the canonical momentum
equals $(i\hbar )^{-1}$ times the commutator of their
individual quantum representations, which amounts to a natural
extension of standard nonunique canonical quantization recipe.
Indeed this extension also resolves the operator-ordering ambiguity
in accord with the Born-Jordan rule, and provides an alternate route
to the unique quantization embodied in the generalized Kerner-Sutcliffe
formula.  The discussion below is, for notational simplicity, initially
restricted to a single degree of freedom; results pertaining to multiple
degrees of freedom and the quantization of fields are given at the end.

An arbitrary infinitesimal transformation of the classical
canonical variables $(q, p)$ for a single degree of freedom
may be written:
\begin{equation}
q\to Q(q,p) = q + \delta Q(q,p),\;\>
p\to P(q,p) = p + \delta P(q,p).
\label{Eq:ArbitrTrans}
\end{equation}
For this transformation to be canonical, the Poisson bracket
$\{ Q, P\}$ of the transformed variables must be equal to
unity.  To first order of the above infinitesimal
transformation, this implies that
\begin{equation}
{\partial(\delta Q)\over\partial q}\, +\,
{\partial(\delta P)\over\partial p}\, = 0.
\label{Eq:CanonReq}
\end{equation}
Thus $\delta P(q,p)$ is not independent of $\delta Q(q, p)$.
Indeed, in order to satisfy Eq.~(\ref{Eq:CanonReq}) above,
one may express {\em both\/} $\delta P$ and
$\delta Q$ in terms of a {\em single\/}, essentially
{\em arbitrary\/} real function $G(q, p)$ as follows:
\begin{equation}
\delta Q =
(\delta\lambda){\partial G\over\partial p}\> ,\;\>
\delta P =
-(\delta\lambda){\partial G\over\partial q}\> ,
\label{Eq:CanonInftes}
\end{equation}
where $\delta\lambda$ is an infinitesimal parameter whose
dimension is seen to be that of action divided by the
dimension of the essentially arbitrary real-valued
{\em generating function\/} $G(q, p)$.  If we put
Eqs.~(\ref{Eq:CanonInftes}) back into
Eqs.~(\ref{Eq:ArbitrTrans}), we obtain two coupled
differential equations for the $\lambda$-parameterized
continuous {\em trajectory\/} of successive canonical
transformations which are produced by the particular
generating function $G(q, p)$:
\begin{equation}
{dQ(\lambda )\over d\lambda}\, =\,
\left .\partial G\over\partial p\right
|_{q = Q(\lambda ), p = P(\lambda )}\> ,\;\>
{dP(\lambda )\over d\lambda}\, =\,
-\,\left .\partial G\over\partial q\right
|_{q = Q(\lambda ), p = P(\lambda )}\> .
\label{Eq:CanonTraj}
\end{equation}
The above continuous canonical transformation trajectory
equations have the {\em same form\/} as Hamilton's
equations of motion, which they become when the
generating function $G(q, p)$ is the classical
Hamiltonian function $H(q, p)$; in that case the
dimension of the parameter $\lambda$ is that of time.
Just as Hamilton's equations of motion can be obtained
from a stationary action principle, so also can the
coupled equation pair for the one-parameter continuous
classical canonical transformation trajectory of
Eq.~(\ref{Eq:CanonTraj})---with an action that is
{\em completely analogous\/}:
\begin{equation}
S_G(\lambda_2,\lambda_1;[Q, P]) =
\int_{\lambda_1}^{\lambda_2}d\lambda
\left (P(\lambda)\,{dQ(\lambda)\over d\lambda}\,
- G(Q(\lambda), P(\lambda))\right ).
\label{Eq:Action}
\end{equation}
The corresponding {\em quantum\/} canonical transformation
arising from $\lambda$-parameter values between $\lambda_1$
and $\lambda_2$ thus must, in coordinate representation,
be given by the analogous {\em path integral\/}---but with
weight functional
$\exp (iS_G(\lambda_2,\lambda_1;[Q, P])/\hbar )$---as one has
for the quantum time evolution transformation in the
well-known special case that $G(q, p)$ is the classical
Hamiltonian function $H(q, p)$ \cite{Garrod}:
\begin{eqnarray}
& &\langle q_f | U_G(\lambda_2 - \lambda_1) | q_i \rangle =
\int_{q_i}^{q_f} d[Q]\, d[P]\,
\exp \left (iS_G(\lambda_2,\lambda_1;[Q, P])/\hbar\right)
\equiv \lim_{m\to\infty}
\left (1\over 2\pi\hbar\right )^{m + 1}\times\nonumber\\
& &\left [\prod_{n = m + 1}^1
\int_{-\infty}^{\infty}dq_{n - 1}\int_{-\infty}^{\infty}
dp_n\exp\left ({i\over\hbar}
\left [p_n\Delta_n -\,
{\delta\lambda\over\Delta_n}
\int_{q_{n - 1}}^{q_n}dq\,
G(q, p_n)\right ]\right )\right ]\delta (q_0 - q_i),
\label{Eq:PathInteg}
\end{eqnarray}
where $\Delta_n\equiv q_n - q_{n -1}$, $q_{m + 1}\equiv q_f$,
and $\delta\lambda\equiv (\lambda_2 - \lambda_1)/(m + 1)$.
Note the {\em integrated\/} averaging in
Eq.~(\ref{Eq:PathInteg}) of $G(q, p_n)$ over $q$-values
between $q_{n - 1}$ and $q_n$, which follows {\em exactly\/} from
the standard picture of a polygonal
approximating $q$-path that {\em connects\/} the neighboring nodes
$(\lambda_1 + (n - 1)\delta\lambda,\> q_{n - 1})$ and
$(\lambda_1 + n\delta\lambda,\> q_n)$ with the unique straight line
segment joining them \cite{Feynman,Garrod,Kerner}.  Many textbook
treatments of the path integral approximate this technically
crucial {\em integrated\/} $q$-averaging for neighboring nodes very
crudely---or even ignore it altogether \cite{Feynman,Brown,Das}.
Poor approximations to this averaging have been
defended on the ground that the neighboring-node $q$-values
$q_{n - 1}$ and $q_n$ approach each other in the path integral
limit \cite{Cohen70}.  That serious misconception is quickly
dispelled by examination of Eq.~(\ref{Eq:PathInteg}), where
it is seen that these neighboring-node $q$-values are
integrated {\em independently\/} along {\em infinite\/} ranges, no
matter {\em how\/} small $\delta\lambda$ becomes as $m\to\infty$.

Given the complete analogy between the above mathematical
structure describing {\em general\/} one-parameter
continuous canonical transformations and its
well-understood special case that the generating function
$G(q, p)$ becomes the classical Hamiltonian function
$H(q, p)$, we can immediately conclude that the quantum
canonical transformation operator
$U_G(\lambda_2 - \lambda_1)$ is unitary and expressible
in the form
$\exp (-i\widehat G(\lambda_2 - \lambda_1)/\hbar)$,
where $\widehat G$ (in alternate notation,
$(G(q, p))_{\rm{op}}$) is the Hermitian quantum operator
that corresponds to the real-valued classical generating
function $G(q, p)$---which itself is an essentially
{\em arbitrary\/} real classical dynamical variable.
Thus we have that
\begin{eqnarray}
\langle q_f |\widehat G| q_i\rangle\equiv
\langle q_f |(G(q, p))_{\rm{op}}| q_i\rangle \!\! &=&\!\!
i\hbar\, \lim_{\lambda_2\to\lambda_1}\,
{\partial\over\partial\lambda_2}\int_{q_i}^{q_f} d[Q]\, d[P]\,
\exp \left (iS_G(\lambda_2,\lambda_1;[Q, P])/\hbar\right)\nonumber\\
&=&\!\! {1\over 2\pi\hbar}\int_{-\infty}^{\infty}dp\,
{\exp\left (ip\, (q_f - q_i)/\hbar\right )\over q_f - q_i}\,
\int_{q_i}^{q_f}dq\, G(q, p),
\label{Eq:ClosForm}
\end{eqnarray}
a result that holds for {\em any\/} $m = 0, 1, 2,\ldots\,$
of the path integral approximating sequence
of Eq.~(\ref{Eq:PathInteg}), and which makes it manifest
that $\langle q_f |\widehat G| q_i\rangle$ is a
{\em Hermitian} (continuous) matrix.  Eq.~(\ref{Eq:ClosForm})
generalizes the Kerner-Sutcliffe quantization formula for
classical Hamiltonians \cite{Kerner} to arbitrary classical
dynamical variables.  It may be contrasted with similar but
inexact quantization formulas such as the ``mid-point
prescription'' \cite{Das}, which have arisen from crude
nonintegrated approximations to the exact path-integral
neighboring-node $q$-averaging discussed above.  As special cases
of Eq.~(\ref{Eq:ClosForm}), we obtain the familiar results that
\begin{equation}
\langle q_f |\widehat{G_1}(q)| q_i\rangle =
G_1(q_f)\, \delta(q_f - q_i),\;\>
\langle q_f |\widehat{G_2}(p)| q_i\rangle =
{1\over 2\pi\hbar}\int_{-\infty}^{\infty}dp\,
\exp\left (ip\, (q_f - q_i)/\hbar\right )\, G_2(p).
\label{Eq:qpCases}
\end{equation}
{}From Eqs.~(\ref{Eq:qpCases}) we readily calculate the
commutator of $\widehat{G_1}(q)$ with $\widehat{G_2}(p)$:
\begin{eqnarray}
\langle q_f|[\widehat{G_1}(q),\, \widehat{G_2}(p)]|q_i\rangle
\!\! &=&\!\!\int_{-\infty}^{\infty}dq'\left [
\langle q_f |\widehat{G_1}(q)| q'\rangle
\langle q' |\widehat{G_2}(p)| q_i\rangle -
\langle q_f |\widehat{G_2}(p)| q'\rangle
\langle q' |\widehat{G_1}(q)| q_i\rangle\right ]\nonumber\\
&=&\!\!\left (G_1(q_f) - G_1(q_i)\right )
\langle q_f |\widehat{G_2}(p)| q_i\rangle.
\label{Eq:Commutr}
\end{eqnarray}
For the operator representation of the Poisson bracket
of $G_1(q)$ with $G_2(p)$, Eq.~(\ref{Eq:ClosForm})
yields:
\begin{eqnarray}
\langle q_f|(\{ G_1(q),\, G_2(p)\} )_{\rm{op}}|q_i\rangle
\!\! &=&\!\!{1\over 2\pi\hbar}\int_{-\infty}^{\infty}dp\,
{\exp\left (ip\, (q_f - q_i)/\hbar\right )\over q_f - q_i}\,
\int_{q_i}^{q_f}dq\,{dG_1(q)\over dq}\,{dG_2(p)\over dp}\nonumber\\
&=&\!\! -\,{i\over\hbar}\, {G_1(q_f) - G_1(q_i)
\over 2\pi\hbar}\int_{-\infty}^{\infty}dp\,
\exp\left (ip\, (q_f - q_i)/\hbar\right )\, G_2(p).
\label{Eq:PoisBrac}
\end{eqnarray}
Thus, by comparing Eqs.~(\ref{Eq:Commutr}) and
(\ref{Eq:PoisBrac}) and taking note of the second
one of Eqs.~(\ref{Eq:qpCases}), we obtain
\begin{equation}
[\widehat{G_1}(q),\, \widehat{G_2}(p)] =
i\hbar(\{ G_1(q),\, G_2(p)\} )_{\rm{op}}.
\label{Eq:CanonQuant}
\end{equation}
Eq.~(\ref{Eq:CanonQuant}) is a natural extension of
the nonunique standard canonical quantization recipe,
which, in fact, turns out to resolve the latter's
operator-ordering ambiguity.  As a key example of
such resolution of the operator-ordering ambiguity,
one readily obtains from Eq.~(\ref{Eq:ClosForm}) that
\begin{equation}
\widehat {q^lp^m} = {1\over l + 1}\sum_{k = 0}^l
\widehat {q^{l-k}}\>\widehat {p^m}\>\widehat {q^k} =
{1\over l + 1}\sum_{k = 0}^l
{\widehat q}\,{}^{l-k}\>{\widehat p}\,{}^m\>{\widehat q}\,{}^k,
\label{Eq:Monomial}
\end{equation}
namely the Born-Jordan ordering rule \cite{Born,Cohen66},
which {\em as well\/} follows (by induction) from the
standard canonical quantization recipe for just the
commutation relations between $\widehat q$ and $\widehat p$,
{\em provided\/} one first applies the extended
canonical quantization principle of
Eq.~(\ref{Eq:CanonQuant}) to obtain
\[
\widehat {q^lp^m} = [(l + 1)(m +1)]^{-1}
(\{ q^{l + 1},\, p^{m + 1}\} )_{\rm{op}} =
[i\hbar (l + 1)(m + 1)]^{-1}
[{\widehat q}\,{}^{l + 1},\> {\widehat p}\,{}^{m + 1}].
\]
Since arbitrary classical dynamical variables $G(q, p)$
may be approximated by polynomials in $q$ and $p$, the
fact that Eq.~(\ref{Eq:CanonQuant}) yields the same
result as Eq.~(\ref{Eq:ClosForm}) for all such
polynomials shows the equivalence of the natural
extended canonical quantization principle of
Eq.~(\ref{Eq:CanonQuant}) to the generalized
Kerner-Sutcliffe quantization formula of
Eq.~(\ref{Eq:ClosForm}).

For multiple degrees of freedom, it is readily
shown that Eq.~(\ref{Eq:ClosForm}) becomes
(see also Ref.~\cite{Kerner}):
\begin{equation}
\langle\vec{q}_f|\widehat G|\vec{q}_i\rangle =
\int {d^n\vec{p}\over (2\pi\hbar)^n}\>
\exp(i\vec{p}\cdot(\vec{q}_f - \vec{q}_i)/\hbar)\,
\int_{-1/2}^{1/2}dw\; G((\vec{q}_f + \vec{q}_i)/2 +
w(\vec{q}_f - \vec{q}_i),\> \vec{p}),
\label{Eq:VClosForm}
\end{equation}
from which one readily derives that the extended canonical
quantization principle of Eq.~(\ref{Eq:CanonQuant}) becomes
\begin{equation}
[\widehat{G_1}(\vec{q}),\,\widehat{G_2}(\vec{p})] =
i\hbar(\{ G_1(\vec{q}),\, G_2(\vec{p})\} )_{\rm{op}} =
i\hbar(\nabla_{\vec{q}}\, G_1\cdot
\nabla_{\vec{p}}\, G_2)_{\rm{op}}.
\label{Eq:VCanonQuant}
\end{equation}
It is to be noted that the quantization of products of
classical dynamical variables do not always factorize as
one might naively expect.  For example, from either
Eq.~(\ref{Eq:VClosForm}) or Eq.~(\ref{Eq:VCanonQuant})
it can be shown that $(q_1p_1q_2p_2)_{\rm{op}} =
(q_1p_1)_{\rm{op}}\, (q_2p_2)_{\rm{op}} - \hbar^2/12$,
where, of course, $(q_ip_i)_{\rm{op}} =
(\widehat{q_i}\widehat{p_i} + \widehat{p_i}\widehat{q_i})/2$
for $i = 1,2$.

Finally, given a set of $m$ real-valued classical dynamical
fields $\phi_k(\vec{x})$ and their $m$ respective canonically
conjugate momentum densities $\pi_k(\vec{x})$, where
$k = 1,\ldots ,m$, the usual canonical quantization recipe
\begin{equation}
[\widehat{\phi_j\mathstrut}(\vec{x}),\,
\widehat{\pi_k\mathstrut}(\vec{y})] =
i\hbar\delta_{jk}\,\delta^{(n)}(\vec{x} - \vec{y})
\label{Eq:FCanonRecp}
\end{equation}
is extended by analogy with Eq.~(\ref{Eq:VCanonQuant}) to
become the canonical quantization principle
\begin{equation}
\left [\widehat{G_1}[\phi_1,\ldots ,\phi_m],\,
\widehat{G_2}[\pi_1,\ldots ,\pi_m]\right ] =
i\hbar\left (\sum_{k =1}^m\int d^n\vec{x}\;\>
{\delta G_1\over\delta\phi_k}\,{\delta G_2\over\delta\pi_k}
\right )\!\! {}_{\rm{op}},
\label{Eq:FCanonQuant}
\end{equation}
where $G_1[\phi_1,\ldots ,\phi_m]$ and $G_2[\pi_1,\ldots ,\pi_m]$
are essentially arbitrary {\em functionals\/} of their respective
field arguments, while $\delta G_1/\delta\phi_k$ and
$\delta G_2/\delta\pi_k$, where $k = 1,\ldots ,m$, denote their
respective partial functional derivatives.

\subsection*{Acknowledgment}

The author wishes to thank T. Sen for discussions and
literature references.

\newpage

%              REFERENCES

\end{document}